\documentclass[11pt,a4paper]{article} 
\usepackage{jheppub}
\usepackage{mciteplus}

\def\gs{\mathrel{
   \rlap{\raise 0.511ex \hbox{$>$}}{\lower 0.511ex \hbox{$\sim$}}}}
\def\ls{\mathrel{
   \rlap{\raise 0.511ex \hbox{$<$}}{\lower 0.511ex \hbox{$\sim$}}}}

\newcommand{\be}{\begin{eqnarray}}
\newcommand{\ee}{\end{eqnarray}}

\def\nue{{\nu_e}}

\def\numu{{\nu_{\mu}}}

\def\nutau{{\nu_{\tau}}}

\def\gsim{\:\raisebox{-0.5ex}{$\stackrel{\textstyle>}{\sim}$}\:}

\def\be{\begin{equation}}
\def\ee{\end{equation}}
\newcommand{\ba}{\begin{array}{c}}
\newcommand{\baz}{\begin{array}{cc}}
\newcommand{\bad}{\begin{array}{ccc}}
\newcommand{\bav}{\begin{array}{cccc}}
\newcommand{\baf}{\begin{array}{ccccc}}
\newcommand{\bena}{\begin{eqnarray}}
\newcommand{\eena}{\end{eqnarray}}
\newcommand{\bea}{\begin{equation} \begin{array}{c}}
\newcommand{\eea}{ \end{array} \end{equation}}
\newcommand{\ea}{\end{array}}

\begin{document}

\subheader{\footnotesize\sc Preprint number: DFTT 13/2011}
\title{Downward--going tau neutrinos as a new prospect of detecting dark matter}

\author[a]{N. Fornengo}
\author[a]{V. Niro}

\affiliation[a]{Dipartimento di Fisica Teorica, Universit\`a di Torino and INFN, 
Sezione di Torino \\ via P. Giuria 1, I-10125 Torino, Italy}

\emailAdd{fornengo@to.infn.it}
\emailAdd{niro@to.infn.it}

\abstract{Dark matter trapped in the Sun produces a flux of all flavors
of neutrinos, which then reach the Earth after propagating out of the Sun and
oscillating from the production point to the detector. The typical signal
which is looked at refers to the muon neutrino component and 
consists of a flux of up--going muons in a neutrino detector, the major
source of background being atmospheric neutrinos.
We propose instead a novel signature, namely the possibility of looking at the tau 
neutrino component of the dark matter signal, which is almost background--free in the downward--going
direction, since the tau neutrino amount in atmospheric neutrinos is negligible
and in the down--going baseline atmospheric muon--neutrinos have no time to sizably 
oscillate. 
We analyze the prospects of studying the downward--going tau neutrinos from dark matter
annihilation (or decay) in the Sun in Cherenkov detectors, by looking at hadronic showers
produced in the charged--current tau neutrino interactions and subsequent tau decay.
We discuss the various sources of background (namely the small tau neutrino component in atmospheric neutrinos, 
both from direct production and from oscillations; tau neutrinos from solar corona interactions; 
the galactic tau neutrino component) as well as sources of background due to misidentification of electron and muon events. 
We find that the
downward--going tau neutrinos signal has potentially very good prospects for Mton scale
Cherenkov detectors, the main limitation being the level of misidentification of
non--tau events, which need to be kept at level of percent. 
Several tens of
events per year (depending on the dark matter mass and annihilation/decay channel) are potentially collectible 
with a Mton scale detector, and a 5~$\sigma$ significance discovery is potentially reachable
for dark matter masses in the range from 20 to 300 GeV with a few years of exposure
on a Mton detector.
}

\keywords{ Solar and Atmospheric Neutrinos, Neutrino Physics, Cosmology of Theories beyond the SM, 
Beyond Standard Model}

\maketitle

\section{\label{sec:intro} Introduction}

It is well established by means of astrophysical and cosmological probes
that almost 23\% of the matter present in the Universe is in the form of a non--luminous component, the so--called 
Dark Matter (DM). Although the evidence for
DM in cosmic structures is of gravitational origin, it
is expected that DM particles may produce a large variety of direct or indirect signals,
which represent the subject of a large number of experimental and theoretical studies.
While direct searches exploit the direct scattering of DM particles on the nuclei
of a low--background detector, indirect signals look for the products of DM 
self--annihilation (or decay) occurring in the galactic halo, in the extragalactic
environment or in those celestial bodies, like the Sun and the Earth, where they
can be gravitationally trapped and accumulated before starting the annihilation
(or decay) process. All annihilation (or decay) products are absorbed by the bodies,
except neutrinos: a flux of high--energy neutrinos, in general composed by all three flavours,
can then emerge from the Sun or from the core of the Earth as a signal of the presence of the trapped DM.

The typical way of looking for a signal from DM annihilation in the Sun or in
the Earth is to search for an excess of upward--going muons produced by
charged current interactions of the $\nu_\mu$ flux produced by DM annihilation. The advantage to look at the upward--going neutrino flux though its muon
conversion is represented by the large
conversion area offered by the rock below the detector, and by the well--established
and efficient experimental techniques to detect high energy muons. The main source
of background is therefore represented by the $\numu$--component of atmospheric
neutrinos.

A great deal of work has been devoted to these studies, see for 
instance Refs.~\cite{Gaisser:1986ha,*Silk:1985ax,*Freese:1985qw,*Krauss:1985ks,*Krauss:1985aaa}, and 
Refs.~\cite{Bergstrom:1996kp,*Bottino:1991dy,*Bottino:1994xp,*Bottino:2004qi} for applications 
to the case of neutralino DM. 
Limits on the flux of through--going muons from DM annihilation inside the Sun, the 
Earth and at the Galactic Center have been set using data 
from water Cherenkov detectors like Super--Kamiokande (SK) \cite{Desai:2004pq} and from neutrino telescopes like 
AMANDA and IceCube \cite{Ackermann:2005fr,*Abbasi:2009uz}. 
Competitive limits can also be obtained considering the stopping muons, which are
usually the dominant signal for low--mass DM particles. 
This analysis has been carried out in the context of a light--neutralino signal at SK in Ref.~\cite{Niro:2009mw} 
and in a more general framework in Ref.~\cite{Kappl:2011kz}, where 
  fully contained events were also considered. Recently, it has been discussed in Ref.~\cite{Lee:2011nt} 
the ability of studying the low mass DM range using track events in the DeepCore array 
of the IceCube detector, that can reach an energy threshold as low as 10~GeV.
Detectors designed for other physics searches, such as liquid argon and magnetized 
iron calorimeter detectors, are able to achieve a much better energy and directional 
resolution than water Cherenkov detectors and thus could be used for detecting 
neutrinos from DM annihilations inside the Sun~\cite{Agarwalla:2011yy,*Mena:2007ty}. 
Moreover, the possibility of detecting DM through electron neutrinos in liquid 
scintillation experiments like KamLAND has also been considered, see e.g. Ref.~\cite{Kumar:2011hi}.

In this paper we instead intend to propose and study the feasibility of a different
signature: downward--going tau neutrinos produced by DM annihilation in the Sun. 
The basic motivation is the following: while DM annihilation typically produces similar
amounts of all neutrino flavors, atmospheric neutrinos are largely dominated by the $\nue$ and $\numu$ components, the $\nutau$'s being very suppressed. Neutrino
oscillations transform a fraction of the original atmospheric $\numu$ into $\nutau$,
but this phenomenon is basically inoperative in the down--going direction, due to the much larger
oscillation length (of the order of the Earth diameter) as compared to the average
production height of atmospheric neutrinos, that for definiteness we fixed to be 15~km in the vertical direction (see e.g. 
Ref.~\cite{Honda:2004yz} for a more precise discussion of the production height). 
This fact implies that in the downward direction 
the $\nutau$ produced inside the Sun by DM annihilation (or decay) represents
an almost background--free signal. 

We wish to recall that the $\nutau$ component in the DM neutrino signal coming from the Sun is basically unavoidable: i) the mechanism of production of DM annihilation is hardly flavor--sensitive to the level of producing only $\nue$ and $\numu$; ii) neutrino
oscillations on the baseline of the Sun--Earth distance in any case average
out any production of $\numu$ into a fraction of $\nutau$ able to reach the Earth. 
This implies that
the amount of the DM--signal $\nutau$'s reaching the Earth from the Sun are of the same
order of magnitude as the DM--signal $\numu$'s, but they are basically background--free.
Therefore
they may be a competitive discovery channel as compared to the standard up--going
$\numu$'s signal, which is affected by a much larger background represented by
atmospheric $\numu$'s. 

The background sources for the $\nutau$ channel are represented by the very small
intrinsic $\nutau$ production from cosmic--rays interactions in the atmosphere 
(atmospheric $\nutau$'s) or by residual atmospheric $\numu$ oscillation,  negligible
for down--going but more relevant for almost horizontal fluxes, by oscillation of $\nue$ and $\numu$ 
produced by cosmic--rays interactions
in the solar corona (which represent an irreducible background for our DM signal, since it
comes from the same source, the Sun) and by a galactic plane $\nutau$ flux, mainly given 
by oscillation of $\numu$ produced in cosmic--ray interactions with the interstellar 
medium (this source of background is, in
principle, reducible by angular cuts on the galactic plane position). We will show
that these sources of backgrounds are, by themselves, not posing significant limits
to the DM signal, which therefore represents a potentially viable novel possibility.

We will instead show that the limiting factor for an analysis of the downward--going 
$\nutau$'s is the
ability of a neutrino detector to identify the tau neutrinos. We will concentrate
on water Cherenkov detectors, in order to be able to access the relatively low energies
which are relevant for DM studies ($E_\nu <$ TeV) (in the case of DM with mass from few tens of GeV to few TeV). 
Since next generation apparata of Mton scale are currently
under study, our predictions are particularly suited for future water Cherenkov detectors
like the Hyper--Kamiokande (HK) project~\cite{Nakamura:2003hk}. 

A signal from tau neutrinos, specifically in connection with DM searches, have been also studied in 
Ref.~\cite{Covi:2008jy}, where the possibility to detect $\nu_\tau$ from gravitino DM decay 
in the halo of our Galaxy was discussed.
High energetic neutrinos ($E_\nu\geq$~TeV) coming from astrophysical sources have been 
considered in Refs.~\cite{Jones:2004ir,Dutta:2000jv}, where the flux of $\nu_\tau$ arising 
from neutrino oscillation was discussed. 
The ability to detect atmospheric tau neutrino events through an ultra--large liquid argon detector is described 
in Ref.~\cite{Conrad:2010mh}. An experimental measurement to detect tau atmospheric neutrinos 
by measuring the energy spectra of neutrino induced showers has been suggested in 
Ref.~\cite{Stanev:1999ki}.

The scheme of the present paper is the following. In Sect.~\ref{sec:flux} we will discuss the 
neutrino fluxes coming from DM annihilation and the background fluxes: 
the atmospheric, solar corona and galactic neutrinos. The class of signals relevant
to water Cherenkov detectors and a discussion about the experimental limits in detecting 
$\nu_\tau$ in those type of detectors is given in Sect.~\ref{sec:Exp}. The calculation of the hadronic contained 
events is then presented in Sect.~\ref{sec:Event}, both for the signal and for the various
sources of background. The capabilities of Mton--scale water Cherenkov detectors on the discovery
of the $\nutau$ signal from DM annihilation coming from the Sun is derived in 
Sect.~\ref{sec:SS}, where we discuss also the level of 
sensitivity that can be achieved from this detection channel on the DM--proton scattering cross 
section $\sigma_p$. Conclusions are given in Sect.~\ref{sec:conclusions}.

\section{\label{sec:flux} Neutrino fluxes}

Let us start by discussing the relevant neutrino fluxes produced by DM
annihilation in the Sun. The decay case is easily obtained along
the same line of reasoning. We are interested not only on the source
spectra, but also in the propagated fluxes which reach the Earth, and
that go through both energy redistribution inside the solar medium and oscillation
processes. We will discuss and treat all these phenomena as in Ref.~\cite{Cirelli:2005gh}.

In addition to the DM signal, we need to discuss the relevant sources of background for the $\nutau$ signal, which 
arise from atmospheric, solar corona and galactic cosmic--rays
interactions.

\subsection{\label{sec:fluxDM} Dark Matter}

The neutrino flux at the detector, coming from DM annihilation inside the Sun and
for each of the three neutrino flavours, is defined as:

\be
\frac{d \phi_\nu}{d E_\nu}= \sum_f BR_f \frac{\Gamma_\odot}{4 \pi d^2} \frac{d N^f_\nu}{d E_\nu}
\ee
where $BR_f$ is the DM branching ratio into the annihilation final--state channel $f$,
$d N^f_\nu/d E_\nu$ are the neutrino spectra for each channel $f$, $\Gamma_\odot$ denotes the annihilation rate 
inside the Sun and $d$ is the distance between the Earth and the Sun. 
For definiteness, in our analysis we will consider two specific annihilation channels: i) direct
annihilation into neutrinos with flavor--blind branching ratios, i.e., 
$BR_{\nu_e \bar{\nu}_e} = BR_{\nu_\mu \bar{\nu}_\mu} = BR_{\nu_\tau \bar{\nu}_\tau} = 1/3$;
ii) annihilation into $\tau \bar{\tau}$ leptons with $BR_{\tau \bar{\tau}}=1$.
More general situations are easily implemented.
The DM fluxes $d N^f_\nu/d E_\nu$ are calculated as described in Refs.~ \cite{Cirelli:2005gh,Blennow:2007tw}, considering 
all physical processes that can occur to the neutrinos
from the production point inside the Sun to the detector at the Earth: vacuum and matter neutrino oscillations, neutral 
current and charged current interactions on the Sun's medium. 
We have fixed the neutrino mixing angles as \cite{Schwetz:2011qt}: $\sin^2 \theta_{23}=0.5$, $\sin^2\theta_{12}=0.304$ and $\theta_{13}=0$. 
The annihilation rate $\Gamma_\odot$ is calculated as in 
Ref.~\cite{Gould:1987ir,*Gould:1987ww,*Gould:1991ApJ}. For the DM velocity distribution in the
rest frame of the Galaxy we have assumed a Maxwellian distribution with $\bar{v}=220$~km~s$^{-1}$ and we have fixed 
the local DM density $\rho_0$ to be 0.3~GeV~cm$^{-3}$. The
annihilation rate depends on the DM--proton scattering cross section, which we will fix
to benchmark values in our analysis, as specified in the following.

\begin{figure}[t]
\includegraphics[width=0.44\textwidth]{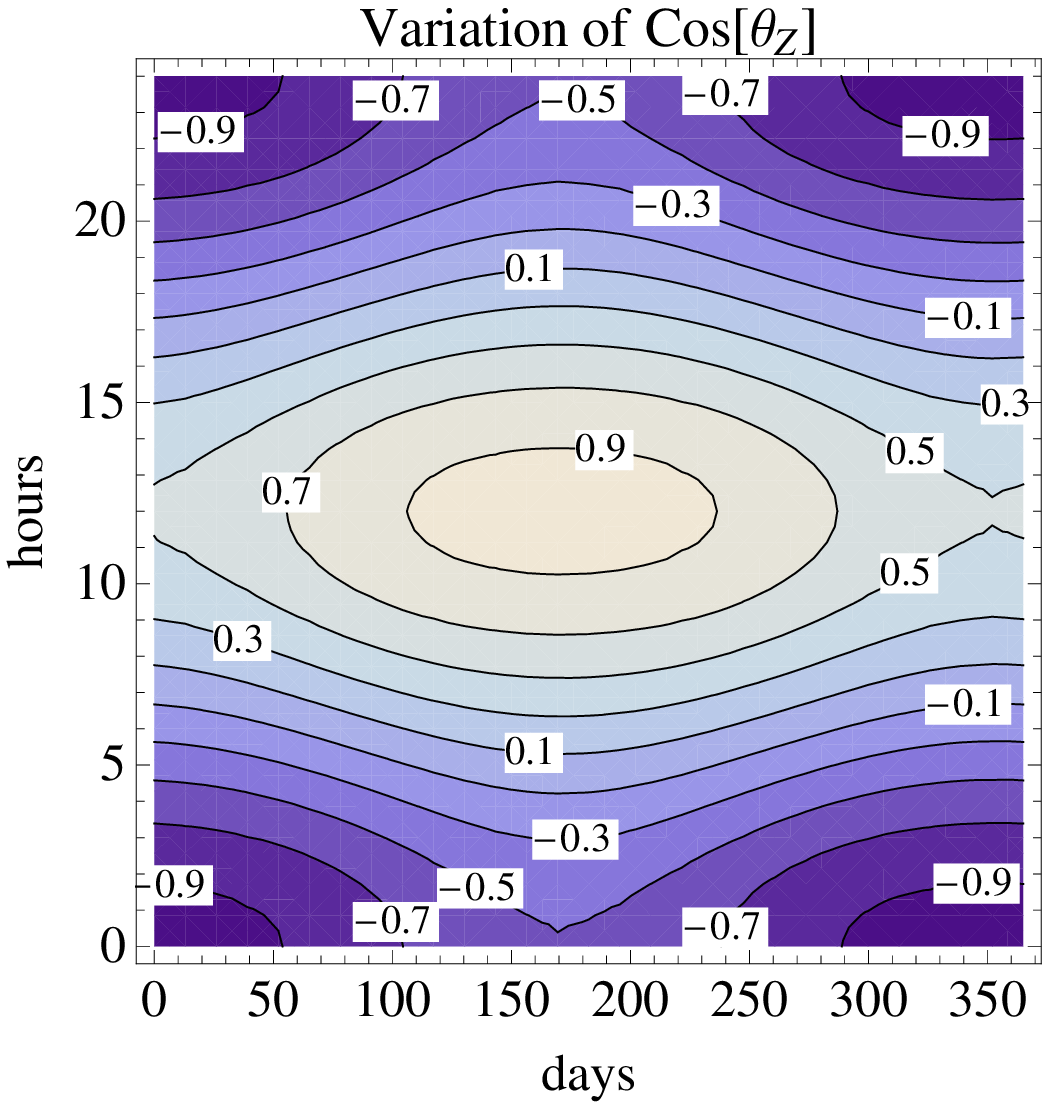}
\includegraphics[width=0.49\textwidth]{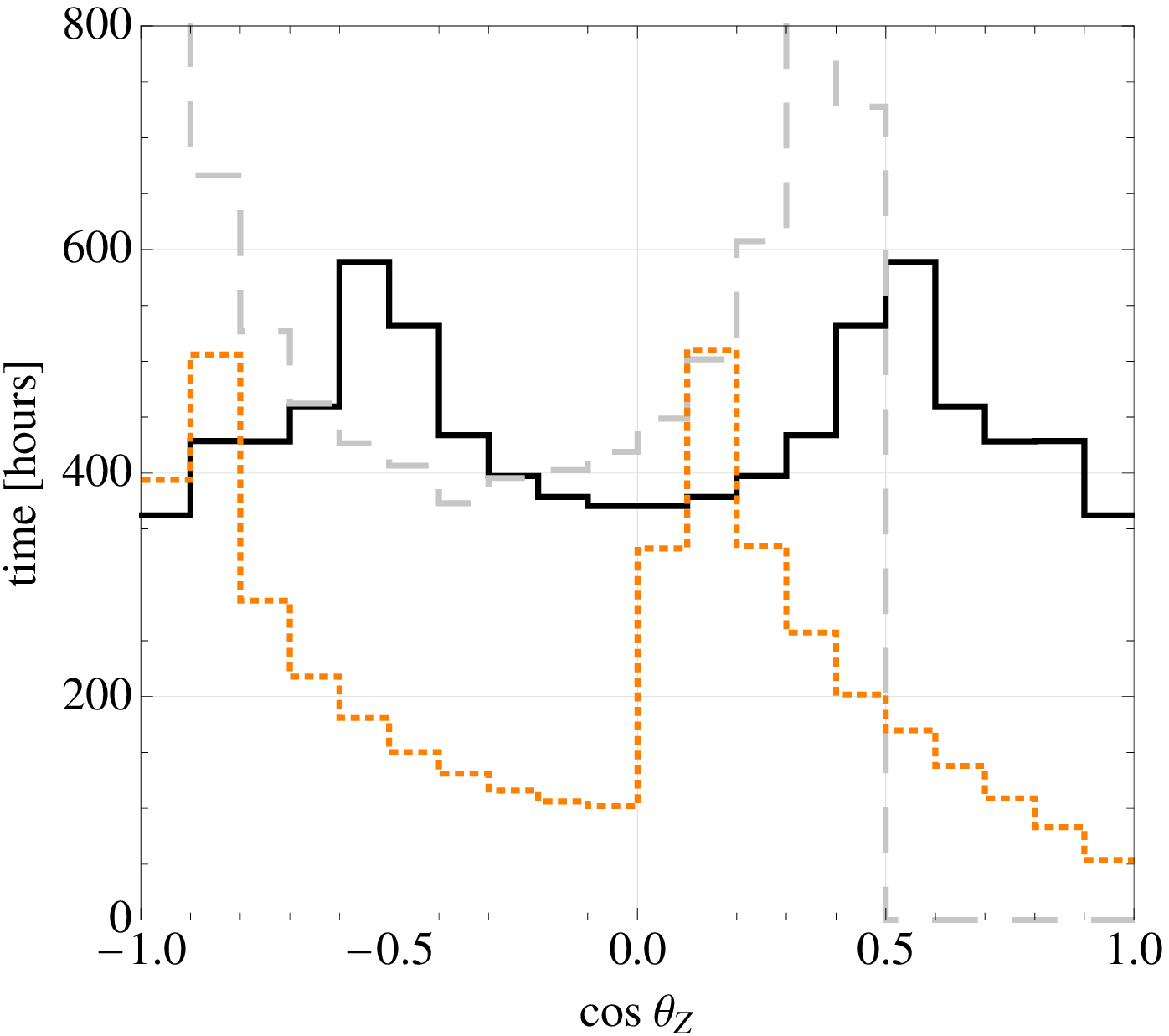}
\caption
{\label{fig:theta_Sun} 
{\sc Left panel:} Apparent motion of the Sun as a 
function of time (hour and day of the year), for a detector located at the latitude $\varphi=36^\circ$. 
The regions denoted by different shades denote the value of the cosine
of the zenith angle (shown by the numbers in labels) occupied by the Sun at a given day and hour.
Positive values of $\cos\theta_Z$ refers to the Sun above the horizon.
{\sc Right panel:} Amount of time (in hours) spent by the Sun in each zenith--angle bin in 
one year, for a latitude $\varphi=36^\circ$ (solid line). The dashed gray 
line shows the time spent by the galactic center in each 
zenith--angle bin. The dotted orange line represents the averaged duty factor of the 
galactic plane convoluted with the position of the Sun, see text for more 
details.}
\end{figure}

Since we are interested in the (downward going) neutrino flux coming from the Sun, it can be important to know the amount 
of time spent by the Sun at different zenith angles above the horizon. This is relevant to determine the duty factor of 
the signal and could also
be exploited for optimizing the signal--to--background ratio, in the case the detection of the 
$\nutau$ flux could be correlated to the Sun's position in the sky. If we 
consider a detector located at a latitude $\varphi$, then the 
motion of the Sun is described by the following expression \cite{Probst:2002}: 

\be
\cos\theta_Z\,=\,\sin\varphi \, \sin \delta \,+\, \cos \varphi \, \cos \delta \, \cos \theta_{\rm HA}
\ee
where $\theta_Z$ is the zenith angle, $\delta$ is the solar declination and $\theta_{\rm HA}$ is the hour angle. 
The value of $\delta$ can be obtained using an approximate expression,
\be
\delta= 23.45^\circ \sin\left( \frac{360^\circ (N_{\rm day}-80)}{365} \right)\,,
\ee
where $N_{\rm day}$ denotes the number of the day, starting on the first of January. 
The hour angle $\theta_{\rm HA}$ is zero at the local solar noon and indicates the time that has 
passed since the Sun was at the local meridian. 
In the left panel of Fig.~\ref{fig:theta_Sun} we report the apparent motion of the 
Sun, considering a detector located at a latitude $\varphi=36^\circ$, that is 
roughly the latitude of the Kamioka site.
In the right panel of Fig.~\ref{fig:theta_Sun} the amount of time (in hours) spent in
one year by the Sun in each zenith--angle bin is shown, for the same latitude: this determines the 
yearly duty--factor for the signal. 
Since $\nutau$'s produced in the galactic plane are one
of our sources of background, in the same right panel of Fig.~\ref{fig:theta_Sun} we also show
the time that the galactic plane spends in each zenith angle bin. 
We have averaged over 120 points in the galactic plane ($b=0$) and 
for each bin above (below) the horizon we have imposed the condition 
that also the Sun position is above (below) the horizon, even if not in the 
same angular bin of the galactic plane. This is done to determine the duty--factor
of the galactic--plane contribution correlated with the presence of the Sun in the
upper (or lower) hemisphere. For this reason, the integral of the galactic plane duty--factor over the 
total zenith angle bins does not turn out to be one year. 
Just for definiteness, we also show in Fig.~\ref{fig:theta_Sun} the duty--factor of the galactic center 
without considering the position of the Sun. In this case the duty--factor is zero for $\cos\theta_Z > 0.5$.

\subsection{\label{sec:fluxBKG} Backgrounds}

There are various sources of background for the $\nu_\tau$ signal 
from DM annihilation. First of all we have the $\nu_\tau$ atmospheric background 
coming from oscillation of atmospheric $\nu_\mu$. This form of background is 
the dominant one for upward--going neutrinos, but it is extremely reduced for 
the downward--going case that we are going to consider. 
For the atmospheric neutrino, we have used the Honda fluxes~\cite{Honda:2011nf}, considering 
the new release from February 2011, where the implementation of the JAM 
interaction model~\cite{Niita:2006zz} for the low energy interactions of cosmic rays and 
air nuclei was implemented. We considered the azimuth--angle--averaged flux at Kamioka site with mountain 
over the detector. 

A second form of background is given by the intrinsic $\nu_\tau$ contribution to 
the atmospheric flux coming from decay of charmed particles. 
This has been computed in Ref.~\cite{Pasquali:1998xf} 
and can be described by the following parameterization:

\be
\log_{10} \left[ E_\nu^3 \frac{d \phi_\nu}{d E_\nu}/
\left( \frac{{\rm GeV}^2}{{\rm cm}^2~{\rm s}~{\rm sr}}\right) \right] 
= - A + B x - C x^2 - D x^3 \,,
\ee
which is valid for $10^2~{\rm GeV} \le E_\nu \le 10^6$~GeV, 
with $x = \log_{10} (E_\nu[{\rm GeV}])$, $A$ = 6.69, $B$ = 1.05, $C$ = 0.150 and $D = -0.00820$. 
We have extrapolated the flux down to $E_\nu^{\rm min}=3.5$~GeV, that is the minimal neutrino 
energy required for the production of a tau lepton. 
In order to check our extrapolation, we have compared the fluxes that we obtain in this way with 
the ones reported in Ref.~\cite{Lee:2004zm}. 
We found that for energies $E_{\nu}\le 10$~GeV our results are in well agreement with~\cite{Lee:2004zm}, 
since the $\nu_\tau$ fluxes are dominated by the oscillated atmospheric neutrinos in that range of 
energies. 
For $10~{\rm{GeV}}~< E_\nu \le 10^2$~GeV, our results are equivalent to~\cite{Lee:2004zm} for the horizontal direction, 
while they differ by at most a factor of two for the down-going direction ($\cos \theta_z=1$). 
We have checked, however, that the events coming from this energy range account for 
roughly the 11\% of the total $\nu_\tau$ events, that we will present in Sect.~\ref{sec:Exp}. 
In the same Section, we will see that the actual experimental background is highly dominated 
by the misidentified events. Thus, we expect that our extrapolation of the intrinsic fluxes 
between 10 and $10^2$~GeV does not have a relevant impact on our analysis (we will comment 
more precisely on this in Sect.~\ref{sec:SS}). 
We want, moreover, to stress that the calculation of the intrinsic $\nu_\tau$ component 
for energies $E_\nu>10^2$~GeV suffers from sizable uncertainties. 
In fact, the atmospheric showering parameters are not precisely known~\cite{Costa:2000jw} and 
in the literature different charm production models are present: quark gluon 
string model~\cite{Kaidalov:1985jg}, recombination quark proton model~\cite{Bugaev:1998bi} and 
perturbative QCD~\cite{Gondolo:1995fq}. 
The fluxes that we use in this article are based on a perturbative QCD approach~\cite{Pasquali:1998xf}, 
but depending on the specific models, the predictions for the fluxes might change of up to 
one orders of magnitude. 
We refer to Refs.~\cite{Costa:2001fb,*Costa:2001va,Costa:2000jw,Enberg:2008te} for a detailed discussion on this topic 
and we will comment in Sect.~\ref{sec:SS} on possible implication of this 
uncertainties on DM searches. 
In the following, when we refer to `atmospheric' $\nu_\tau$ flux we will always imply the 
sum of the oscillated fluxes from atmospheric $\nu_\mu$ and the intrinsic contribution.

The third form of background is represented by the neutrino flux produced in the solar corona 
by cosmic--ray collisions. This has been studied in Ref.~\cite{Ingelman:1996mj}, where 
a suitable parameterization of the flux of electron and muon neutrinos ($j=\nu_e,\nu_\mu$) has
been determined:

\be
\frac{d \phi_j}{d E_\nu}= N_0
\frac{(E_\nu[{\rm GeV}])^{-\gamma-1}}{1+A(E_\nu[{\rm GeV}])}\,
({\rm GeV}~{\rm cm}^2~{\rm s})^{-1}\,, 
\ee
which is valid for $10^2~{\rm GeV} \le E_\nu \le 10^6$~GeV. For the numerical values of the coefficients 
$N_0$, $A$ and $\gamma$ we refer to Ref.~\cite{Ingelman:1996mj}. 
Also in this case we have extrapolated the neutrino fluxes down to $E_\nu^{\rm min}=3.5$~GeV.
During their propagation to the Earth, 
the electron and muon neutrinos produced in the solar corona oscillate and generate $\nu_\tau$. 

Finally, the fourth possible source of background is given by the fluxes of tau neutrinos 
from the Galactic plane. A complete discussion of this topic is presented in 
Ref.~\cite{Athar:2004uk,Athar:2004um}. The $\nu_\tau$ flux can be parameterized as:

\be
\frac{d \phi_\nu}{d E_\nu}= 
9 \times 10^{-6} ({\rm GeV}~{\rm cm}^2~{\rm s}~{\rm sr})^{-1} (E_\nu[{\rm GeV}])^{-2.64}\,,
\ee
which is valid in the energy range $1~{\rm GeV}\le E_\nu \le 10^3$~GeV. 
As discussed in connection with Fig.~\ref{fig:theta_Sun},
for a detector located at a latitude $\varphi=36^\circ$ this contribution starts to be sizable
for $\cos \theta_Z \lesssim0.5$. In this range of zenith angles the atmospheric 
and intrinsic fluxes are by far the dominant ones and thus, for the $\nutau$ signal coming from 
the Sun, the galactic neutrinos do not represent a troublesome source of background. 
We have nevertheless included this contribution in the calculation.

\begin{figure}[t]
\centering
\includegraphics[width=0.48\textwidth]{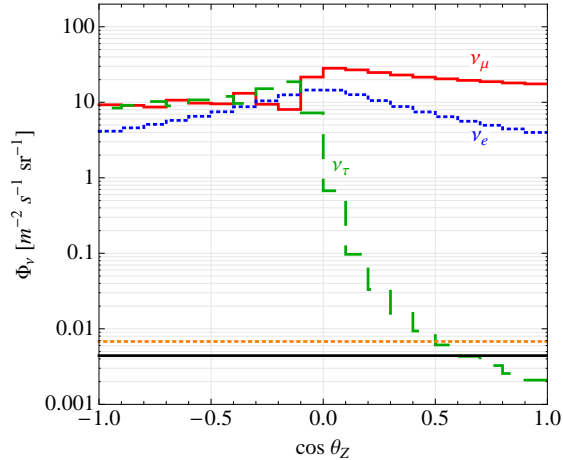}
\caption
{\label{fig:flux_cos} Atmospheric neutrino fluxes as a function of the
zenith angle $\cos \theta_Z$ at Kamioka site. 
The dotted (blue), solid (red) and dashed (green) line refer to the $\nu_e$, $\nu_\mu$
and $\nu_\tau$ fluxes at the detector (with oscillation included).
The solid horizontal (black)
line denotes the $\nu_\tau$ flux from solar corona interaction, 
while the dotted horizontal (orange) line refers to the level of the galactic $\nu_\tau$ flux. 
Note that these two fluxes (differently from the atmospheric contributions) are present 
only when the Sun or the galactic plane occupy each specific angular bin.
Fluxes have been integrated from $E_\nu^{\rm min}=3.5$~GeV.}
\end{figure}

Fig.~\ref{fig:flux_cos} shows the total neutrino fluxes (integrated from $E_\nu^{\rm min}=3.5$~GeV 
up to $E_\nu^{\rm min}=10^4$~GeV) as a function of the zenith angle $\cos \theta_Z$. It is well visible how the 
atmospheric $\nu_\tau$ background is sizebly reduced for $\cos \theta_Z \geq 0$ as compared 
to the $\cos \theta_Z \leq 0$ case and how the $\nu_\tau$ flux is much smaller than 
the $\nu_e$ and $\nu_\mu$ fluxes. This behavior is the main motivation for our proposal
on downward--going tau events. In fact, as already discussed above, the
DM $\nutau$ signal from the Sun is comparable, in size, with the $\numu$ signal. Nevertheless,
the up--going muon channel (measured through up--going muons) has a much larger atmospheric
background than the down--going tau channel, as is clearly seen in Fig.~\ref{fig:flux_cos}.
An up--going muon signal which occurs to be completely dominated by the atmospheric neutrino background, could instead 
dominate the atmospheric background in the down--going tau
channel. In Fig.~\ref{fig:flux_cos} we also show the solar corona and the
galactic $\nutau$ fluxes. 
We notice that solar corona neutrinos become non--negligible mainly 
for $\cos\theta_Z \gsim 0.5$, while the galactic neutrinos are more important 
for angles with $\cos\theta_Z < 0.5$. 
For the down--going $\nutau$ signal, atmospheric neutrinos represent 
the main source of background, while the galactic and solar corona 
neutrinos give a sizable contribution for small and large zenith angles 
respectively. 
In both cases, they are very suppressed, therefore offering potential chances to a signal to emerge.

\begin{figure}[t]
\includegraphics[width=0.48\textwidth]{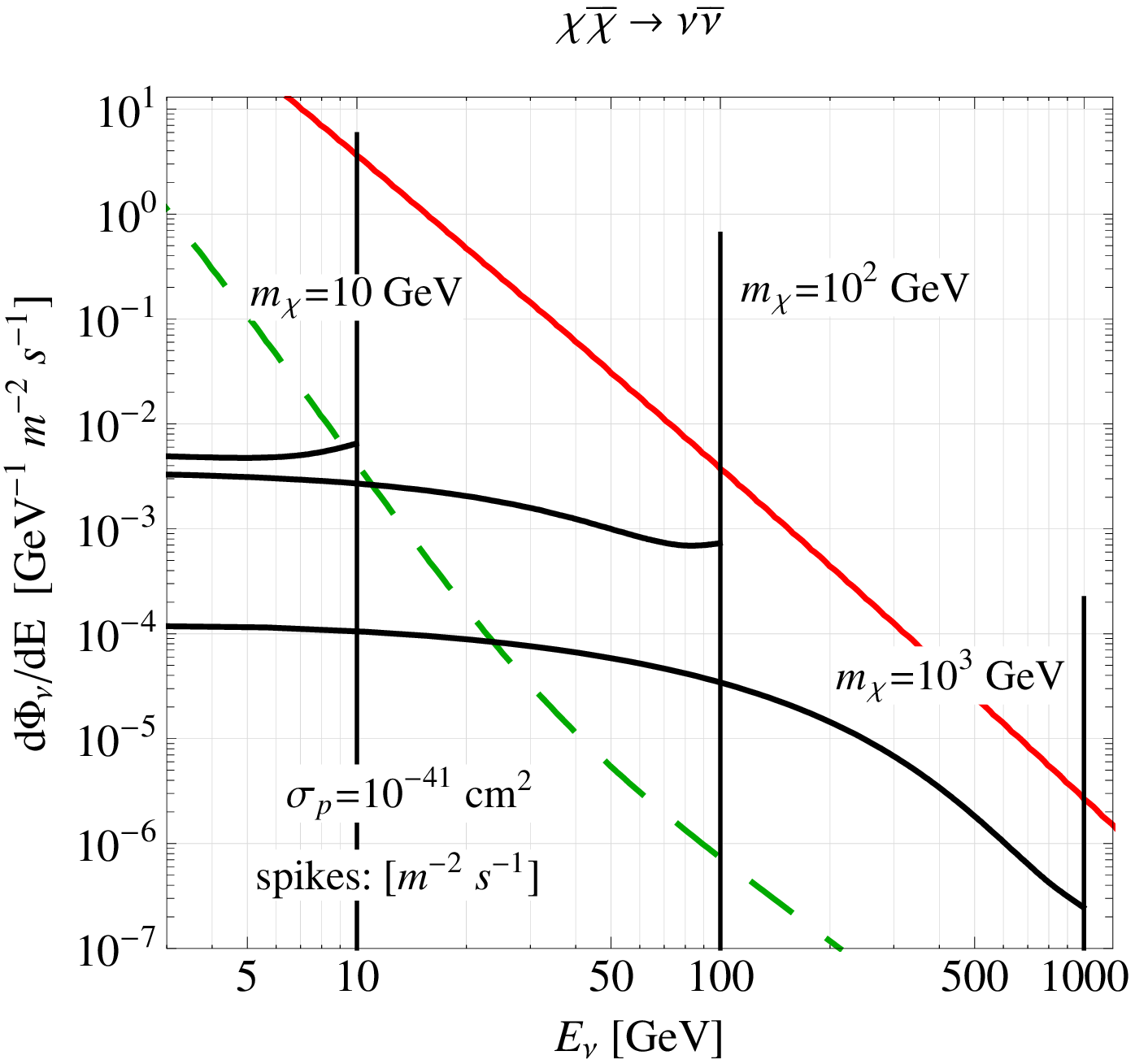}
\includegraphics[width=0.48\textwidth]{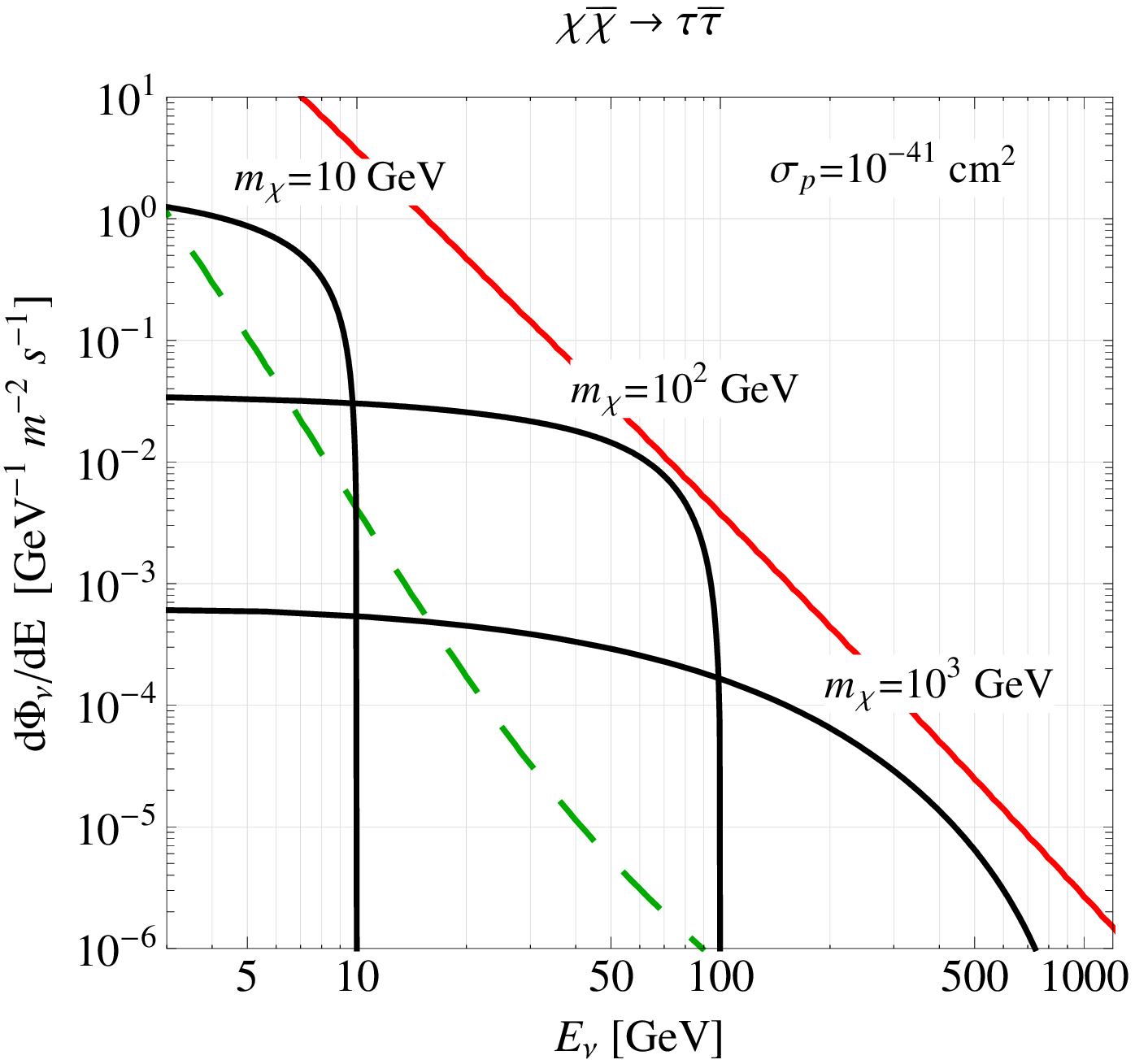}
\caption
{\label{fig:flux} Fluxes of $\nu_\tau$ arising from DM annihilation inside the 
Sun: the left panel refers to the case of direct annihilation into neutrinos 
with even branching ratios in the three flavours, while the right panel
stands for annihilation into tau leptons. 
In the left panel, the continuum part of the spectrum is 
in units of GeV$^{-1}$~m$^{-2}$~s$^{-1}$, while the line is in units of 
m$^{-2}$~s$^{-1}$. 
The DM spin--independent scattering cross section $\sigma_p$ is fixed at
the value 10$^{-41}$~cm$^2$ and the cases for DM masses of 10, 100 and 1000 GeV are reported.
The solid (red) line shows the downward--going flux of atmospheric {\em muon} neutrinos
$\nu_\mu$, 
while the long--dashed (green) line refers to the downward--going background 
flux of $\nu_\tau$ (sum of atmospheric, both intrinsic and from oscillation, 
solar corona and galactic fluxes). The fluxes have been integrated over zenith angles 
with $\cos \theta_Z \geq 0$.}
\end{figure}

A few examples of downward--going $\nu_\tau$ fluxes as a function of the neutrino energy $E_\nu$ 
are shown in Fig.~\ref{fig:flux}, together with the total downward--going $\nu_\tau$ background fluxes: 
sum of atmospheric, both intrinsic and from oscillation, solar corona and galactic fluxes. 
To allow for a realistic comparison between the fluxes, in summing the galactic and solar 
contributions to the atmospheric ones, we have multiplied by the fraction of the time that the 
Sun or galactic plane spend in each zenith angle bin with respect to the total time that the Sun 
spends above the horizon in one year. 
We have integrated over zenith angles with $\cos \theta_Z \geq 0$. 
This represents the angular range that we will consider throughout the paper. 
For the signal, we show the cases for DM mass of 10, 100, 1000 GeV, 
and for the DM spin--independent scattering cross section $\sigma_p$ we have used a benchmark value of 10$^{-41}$~cm$^2$ 
(with $\sigma_p$ we will denote throughout the 
paper the spin--independent scattering cross section, while with $\sigma^{\rm SD}_p$ the 
spin dependent one). The left and right panels refer to the two benchmark annihilation
channels: direct annihilation into neutrinos and annihilation into tau leptons. In the case
of direct annihilation into neutrinos, we clearly notice the line a $E_\nu = m_\chi$
together with the degraded--energy tail due to neutrino propagation in the Sun's interior.
Also from Fig.~\ref{fig:flux} we can see that the signal is significantly dominant over the background in the 
down--going $\nutau$ channel,
especially at high energies where the tau cross section is not suppressed by mass threshold.
 In the same figure, we also show the flux
of down--going atmospheric $\numu$, which (together with the down--going $\nue$)
will produce a source of background events for the down--going 
$\nutau$ signal in water Cherenkov detectors, due to misidentification of the muon and electron events in the 
detector, as will be discussed in Sect.~\ref{sec:Exp}.

\section{\label{sec:Exp} Signals at water Cherenkov detectors}

The experimental signals of high--energetic tau neutrinos have been analyzed in 
Ref.~\cite{DeYoung:2006fg}, where the double bang and lollipop signatures 
were considered. 
These signatures, however, are distinctive only for energies above the PeV range. 
For lower energies, between TeV and PeV range, it might be possible to tag taus which decay to muons, 
if the neutrino interaction vertex occurs within the detector~\cite{DeYoung:2006fg}.

For the range of energies we are interested in (GeV--TeV), 
charged--current $\nu_\tau$ interactions in water Cherenkov detectors will lead to 
multiple Cherenkov rings and the possibility of individually identifying these events 
is currently based only on statistical methods. 
In Ref.~\cite{Abe:2006fu,Kato:2007zz} the SK Collaboration has employed neural network and maximum 
likelihood techniques to successfully discriminate tau neutrino events. They concentrate on the hadronic decays of tau
leptons ($BR_\tau^{h}\simeq 64$\%), since they have a
more spherical topology than the backgrounds. We will exploit the same type of signature
in our calculations.
The primary backgrounds to $\nu_\tau$ charged--current events are neutral--current and 
charged--current events from $\nu_e$ and $\nu_\mu$ atmospheric neutrinos. 
A number of event selection criteria~\cite{Abe:2006fu,Kato:2007zz} can be applied to reduce these backgrounds. 
Using a likelihood analysis and a neural network, the SK Collaboration achieved 
an efficiency (with respect to the total number of events in the fiducial volume) 
of 43.1\% and 39.0\% for tau events identification, respectively.
The electron and muon events are misidentified as tau events with a percentage 
of 3.8\% and 3.4\% for the same two statistical analyses. 
Since the atmospheric $\nu_e$ and $\nu_\mu$ fluxes
are usually larger than or of the same order of the DM fluxes, as can be seen in Fig.~\ref{fig:flux}, 
the misidentification has a relevant impact on the actual performance of 
water Cherenkov detectors in identifying $\nu_\tau$ events.

In our analysis we have focused on the favourable situation in which the track 
events from $\nu_\mu$ charged current interactions will always be detected and 
correctly identified. For this reason, in the misidentified background events 
we will only considered neutral--current events from $\nu_e$ and $\nu_\mu$ atmospheric 
neutrinos and charged--current events from $\nu_e$ interactions. 
Note that in principle also $\nu_e$ and $\nu_\mu$ from solar corona interactions 
will contribute to this form of background, but they are negligible with respect to 
the atmospheric $\nu_e$ and $\nu_\mu$. 
Another possible form of background events is represented by neutral--current 
events from $\nu_\tau$ (atmospheric, intrinsic and solar corona contributions) and 
by $\nu_\tau$ charged--current interaction, with tau decaying into electron. 
We will neglect these events since we will numerically see that the background is 
typically dominated by neutral current $\nu_e$ and $\nu_\mu$ and by charged-current $\nu_e$ 
misidentified events. 
Finally, we wish to comment that also $\nu_\mu$ and $\nu_e$ fluxes from DM annihilations
would contribute to the signal events, again through misidentification. We will not include
this type of contribution, which is marginal in determining the amount of signal events.

\section{\label{sec:Event} Contained hadronic events}

The class of signal events we are considering is represented by the charged--current
production of a tau lepton, followed by its hadronic decay. The hadronic showers 
of the decay produce Cherenkov rings in the water
detector. In this Section we discuss our determination of 
the number of hadronic events, and the calculation of the
class of events which contribute to the background through misidentification, namely
background events from $\nue$ and $\numu$ neutral current interactions and from
$\nue$ charged current interactions.

\subsection{\label{sec:EventDM} Signal events from hadronic tau decay}

\begin{figure}[t]
\includegraphics[width=0.96\textwidth]{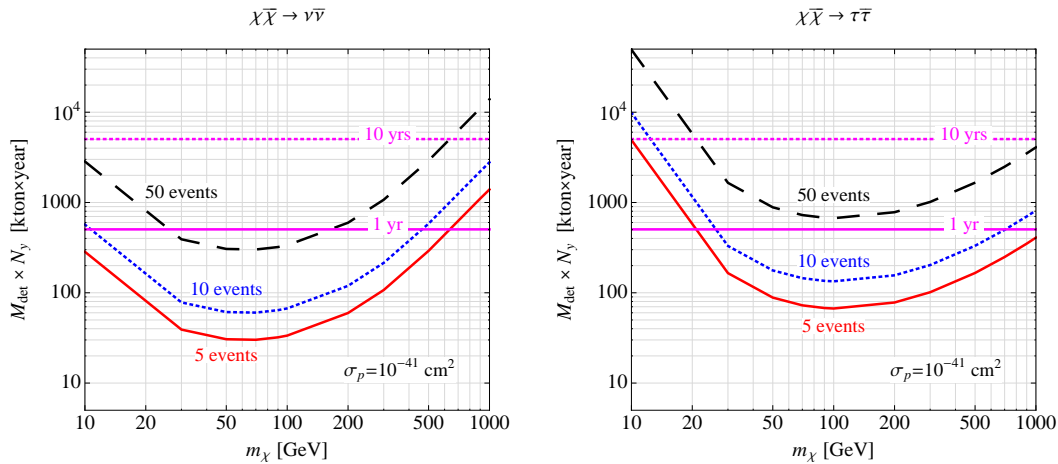}
\caption
{\label{fig:events_exposure} 
Iso--contours of number of downward--going $\nutau$ hadronic events 
in the plane of detector exposure (in kton $\times$ year) and DM
mass $m_\chi$. The left panel refers to DM
annihilation into neutrinos, while the right panels
stands for annihilation into tau leptons. The elastic scattering cross section
on protons (relevant for capture in the Sun) is fixed at the benchmark value of $10^{-41}$ cm$^{2}$.
The solid (red), dotted (blue)
and dashed (black) lines refer to 5, 10 and 50 events in the detector, respectively.
The two horizontal (pink) lines denote the exposures that can be reached
by a 0.5 Mton detector, like HK, in 1 (solid) and 10 (dotted) years.}
\end{figure}

For the calculation of contained tau events, we will use the derivation of 
Ref.~\cite{Giordano:2010pr}, where the concept of visible energy is introduced. On this
topic, see also Refs.~\cite{Nunokawa:2003ep,Dutta:2000jv,Stanev:1999ki}. 
For this category of events, the visible energy $E_{\rm vis}$ is the sum of the 
energy $E_{h, 1}$ of the broken nucleon 
and the hadronic energy $E_{h, 2}$ of the tau decay. 
The total number of contained events for charged current 
$\nu_\tau$ interactions is:

\be
\left.N_\tau^{\mathcal{CC}}\right|_\mathcal{S,B}=
M_{\rm det} N_y \times 
\int^{E^{\rm max}_{\rm vis}}_{E^{\rm min}_{\rm vis}}
 dE_{\rm vis}
\int d \Omega\;\; \eta(\theta) \,
\left.\frac{d\mathcal{I}^\mathcal{CC}_\tau}{d \Omega\,d E_{\rm vis}}\right|_{\mathcal{S,B}}\,,
\label{eq:uno}
\ee
where $M_{\rm det}$ is the detector mass, $N_y$ the number of years of exposure and $\eta(\theta)$ 
is the on--source duty factor. For the signal, we use the duty--factor
discussed in Sect.~\ref{sec:fluxDM} and shown in the right panel of Fig.~\ref{fig:theta_Sun}.
For the background events we have conservatively considered the case in which 
the events are classified only as upward/downward. Thus, the background is not
filtered through the duty--factor of Fig.~\ref{fig:theta_Sun}, instead it is always
present when the Sun is above the horizon ($\eta=1/2$). 
In a more optimistic scenario in which the detector would be able to correlate the
direction of the observed Cherenkov rings of the hadronic showers with the
position of the Sun, a reduction of the background would be possible.
We will nevertheless neglect this optimistic possibility here and we will consider only the 
more conservative case in which the events are identified as upward--going or downward--going. 

The function $d\mathcal{I}^\mathcal{CC}_\tau/d \Omega\,d E_{\rm vis}$ in Eq~.\eqref{eq:uno} 
is defined as:

\bena
\left.\frac{d\mathcal{I}^\mathcal{CC}_\tau}{d\Omega\,d E_{\rm
vis}}\right|_{\mathcal{S,B}}&=&
\int dE_\nu
\int dE_\tau
\left.
\frac{d \phi_{\nu_\tau}}{d \Omega d E_\nu}\right|_{\mathcal{S,B}}
\Sigma^{\mathcal{CC}}_\tau(E_\tau,E_\nu)\,
\frac{d \Gamma_h}{d E_{\rm vis}}
+(\nu \rightarrow \bar{\nu})\,,
\label{eq:CCnu1}
\eena
where $\cal S$ and $\cal B$ denote signal and background, respectively,
$d \phi_{\nu_\tau}/d \Omega d E_\nu$ is the $\nu_\tau$ flux coming from
DM annihilation or the background flux of atmospheric, intrinsic and solar
corona tau neutrinos.
The function $\Sigma^{\mathcal{CC}}_\tau$, defined as:
\be
\Sigma^{\mathcal{CC}}_\tau(E_\tau,E_\nu)=N_A\left(
\mathcal{Z} \frac{d \sigma^p_{\nu_\tau}}{d E_\tau}(E_\tau,E_\nu)
+\mathcal{N} \frac{d \sigma^n_{\nu_\tau}}{d E_\tau}(E_\tau,E_\nu)
\right)
\label{eq:sigmaCC}
\ee
quantifies the number of interactions, with $N_A$ being the Avogadro's
number,
$\mathcal{Z}$ and $\mathcal{N}$ the fraction of proton and neutrons
present in
the detector and $d \sigma^{p,n}_{\nu_\tau}/d E_\tau$ the $\nu_\tau$ charged--current cross section on proton 
and neutron, for which we adopt Ref.~\cite{Jeong:2010nt,*Kretzer:2002fr},
where the correction due to the finite tau mass is implemented.
For the parton distribution functions (PDF) we have chosen the MSTW 2008
NNLO~\cite{Martin:2009iq,*LinkPDF}
varying $Q^2$ from 5~{\rm GeV}$^2$ till 10$^4$~{\rm GeV}$^2$. For lower
$Q^2$ we have frozen the
PDF to the values assumed at $Q^2=$5~{\rm GeV}$^2$. For a more refined
calculation,
it should be taken into account also the non-perturbative~\cite{Reno:2006hj}
evolution of the
PDF in the low $Q^2$ region, $Q^2 \lesssim$~2~GeV$^2$.
Moreover, also target mass correction~\cite{Kretzer:2003iu} could have some
impact for very precise estimates. Nevertheless,
we have neglected these two latter corrections, since they are beyond the
precision required for
our study. We have checked that with our assumption we achieve an agreement better
than 10\% with the total cross section reported
in Ref.~\cite{Jeong:2010nt,*Kretzer:2002fr}.

The function $d\Gamma_h/dE_{\rm vis}$ in Eq.~\eqref{eq:CCnu1} is defined as the decay rate 
of tau leptons into hadrons, provided
that the energy of the hadronic decays $E_{h,2}$ is equal to the tau
lepton energy minus the final neutrino
energy $E_{\nu_{f}}$, with $E_{\nu_{f}}=E_\nu-E_{\rm vis}$:
\be
\frac{d \Gamma_h}{d E_{\rm vis}} \equiv \int dE_{h,2} \,\frac{dn_h}{d
E_{h,2}}\,
\delta(E_{h,2}-\left(E_\tau-\left(E_\nu-E_{\rm vis}\right)\right))
\,\Theta(E_\nu-E_{\rm vis})
\ee
The decay rate $dn_h/d E_{h,2}$ can be obtained from the $\nu_\tau$
spectra produced in the
decay:
\be
\frac{d n_h}{d{E_{h,2}}}=
\frac{1}{E_\tau}
\frac{d n_h}{d z_h}\qquad
\text{with}\qquad z_{h}=\frac{E_{h,2}}{E_\tau}\,,
\ee
where
\be
\frac{d n_h}{d z_h}=\frac{d n_\nu}{d z_\nu}\qquad
\text{with}\qquad z_\nu=\frac{E_\nu-E_{\rm vis}}{E_\tau}\,.
\ee
The expressions of $d n_\nu/d z_\nu$ are given in
Refs.~\cite{Pasquali:1998xf,Dutta:2000jv,Lipari:1993hd}. Note that the
function $dn_h/d E_{h,2}$ is normalized to the total branching ratio of
tau into hadrons, $BR^h_\tau=0.64$.

\begin{figure}[t]
\centering
\includegraphics[width=0.48\textwidth]{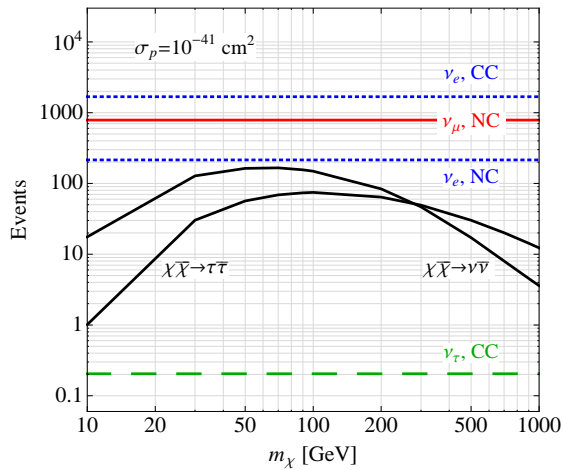}
\caption
{\label{fig:events} Number of downward--going $\nutau$ hadronic events (solid lines)
as a function of the 
DM mass $m_\chi$, for a scattering cross section on proton $\sigma = 10^{-41}$
cm$^2$  and for a detector exposure $M_{\rm det} N_y=$1~Mton$\times$year. 
We show also as horizontal lines 
the neutral current events expected from the atmospheric $\nu_e$ (lower dotted blue line) 
and $\nu_\mu$ (solid red curve), the charged--current events from $\nu_e$ (upper
dotted line) and the charged--current 
events expected from the background $\nu_\tau$.
}
\end{figure}

Using the previous definitions, we can rewrite Eq.~\eqref{eq:CCnu1} as:
\bena
\left.\frac{d\mathcal{I}^\mathcal{CC}_\tau}{d\Omega\,d E_{\rm
vis}}\right|_{\mathcal{S,B}}&=&
\int^{\infty}_{E_{\rm vis}} dE_\nu
\int^{E_\nu}_{E^{\rm min}_\tau} dE_\tau
\left.
\frac{d \phi_{\nu_\tau}}{d \Omega d E_\nu}\right|_{\mathcal{S,B}}
\Sigma^{\mathcal{CC}}_\tau(E_\tau,E_\nu)
\frac{1}{E_\tau}\,\frac{d n_\nu}{d z_\nu}\,,
\eena
with an analogous part for antineutrinos and with 
$E^{\rm min}_\tau={\rm max}[m_\tau,E_\nu-E_{\rm vis}]$. 
We have considered a lower limit on the visible energy 
equal to the minimal neutrino energy for tau lepton production: 
$E^{\rm min}_{\rm vis}=3.5$~GeV. 
A specific detector should choose the minimal visible energy able to 
maximize the number of correctly identified tau events with respect to the misidentified 
ones. This quantity could turn out to be different from our choice of $E^{\rm min}_{\rm vis}$, but 
this would be highly detector dependent and should be obtained through dedicated Monte Carlo 
analyses. 
We fix the upper limit on the visible energy to 
$E^{\rm max}_{\rm vis}=10^4$~GeV for the atmospheric events, or to 
$E^{\rm max}_{\rm vis}=m_\chi$ for the DM events.

The number of signal events expected for a detector of a given exposure (expressed
in kton $\times$ year) is shown in Fig.~\ref{fig:events_exposure} as a function
of the DM mass, for the two cases of DM
annihilation into neutrinos (left panel) and annihilation into tau leptons
(right panel). The elastic scattering cross section
on protons, relevant for capture in the Sun and that determines the size of
the annihilation rate $\Gamma_\odot$, is fixed at the benchmark value of $10^{-41}$ cm$^{2}$. 
In all our analyses we are considering that equilibrium between capture and
annihilation has been reached, as typically occurs for the Sun.
The solid (red), dotted (blue)
and dashed (black) lines refer to 5, 10 and 50 events in the water Cherenkov detector, respectively. 
The two horizontal (pink) lines denote the exposures that can be reached
by a 0.5 Mton detector, like HK, in 1 (solid) and 10 (dotted) years. We notice that
for a Mton--scale detector, the expected number of down--going $\nutau$ signal events
can reach the level of 50 or more, depending on the DM mass. For a prolonged exposure of 
a decade or more, more than a hundred of signal events is potentially
under reach. For the annihilation channels under study, this signal is most sensitive to 
DM masses in the range from 30 GeV up to 200-300 GeV. For lighter DM, the signal
is reduced, since the available energy to be transferred to the hadronic showers
in the tau channel is reduced (we remind that only neutrinos with energy 
$E_\nu \leq m_\chi$
are produced by the non--relativistic annihilation process). For heavier DM the signal
fades away since capture is less efficient and moreover absorption processes of higher
energy neutrinos in the Sun start to become operative \cite{Cirelli:2005gh,Blennow:2007tw}.
These properties can also be observes in Fig.~\ref{fig:events}, where the 
expected
number of downward--going $\nutau$ hadronic events are shown (as solid lines)
as a function of the DM mass $m_\chi$, for the same scattering cross
section on proton of Fig.~\ref{fig:events_exposure}, for a detector exposure $M_{\rm det} N_y=$1~Mton$\times$year. 
In the same figure, we also show the charged--current events expected from the background $\nu_\tau$ (horizontal dashed line).
This figure demonstrates that the specific $\nu_\tau$--induced background is negligible.

\subsection{\label{sec:EventBKG} Background events from $\nue$ and $\numu$ neutral--current interactions}

As we have already mentioned before, a non--negligible source of background is 
represented by the atmospheric $\nu_e$ and $\nu_\mu$ neutral current 
interactions, since a fraction of the Cherenkov ring they produce are misidentified 
as hadronic tau events.  
The rate of these events is defined as:

\be
\left.N^{\mathcal{NC}}_{e,\mu}\right|_\mathcal{B}=M_{\rm det} N_y \times 
\int^{E^{\rm max}_{\rm vis}}_{E^{\rm min}_{\rm vis}}
dE_{\rm vis}\,
\int d \Omega\;\;\eta(\theta)\,
\left.\frac{d\mathcal{I}^\mathcal{NC}_{e,\mu}}{d\Omega\,d E_{\rm vis}}\right|_\mathcal{B} \,,
\ee
where $d\mathcal{I}^\mathcal{NC}_{e,\mu}/d\Omega\,d E_{\rm vis}$~\cite{Giordano:2010pr} is:

\bena
\left. \frac{d\mathcal{I}^\mathcal{NC}_{e,\mu}}{d\Omega\,d E_{\rm vis}}\right|_\mathcal{B}&=& 
\int^{\infty}_{E_{\rm vis}} dE_\nu
\left. \frac{d \phi_{\nu_e,\nu_\mu}}{d \Omega d E_\nu} \right|_\mathcal{B}
\Sigma^{\mathcal{NC}}_{e,\mu}(E_{\rm vis},E_\nu)
+(\nu \rightarrow \bar{\nu})\,.
\eena
The function $\Sigma^{\mathcal{NC}}_{e,\mu}$ is the analogous of Eq.~\eqref{eq:sigmaCC} for neutral 
current interactions, for which we use the deep--inelastic cross section of Ref.~\cite{Strumia:2006db}. 
Note that, for neutral current events, the fraction of energy that is transferred to the cascade 
is given by $y=E_{\rm vis}/E_\nu$, with $E_\nu$ being the initial neutrino 
energy.

The number of neutral current $\nu_e$ and $\nu_\mu$ events for one Mton$\times$year exposure are 
reported in Fig.~\ref{fig:events}, as horizontal lines. 
The lower dotted blue line shows the neutral current events expected from the atmospheric 
$\nu_e$, the solid red curve refers to $\nu_\mu$. Together with the charged--current
$\nue$ events (discussed in the next Section), these class of events
are not directly comparable to the signal events, since they represent a backgound
only when they are misidentified as tau hadronic events. 
Fig.~\ref{fig:events_exposure}
shows that these classes of events pose a problem if they are not controlled at a level better than 
a few percent. This level of misidentification is foreseeable, since
SK reconstruction and analysis algorithms are already able to reduce the misidentification
at the level of less than 10\% \cite{Abe:2006fu,Kato:2007zz}. A misidentification level around 1\%, together with an 
efficiency of reconstruction of hadronic tau--events increased of about 70--80\% would 
bring the
down--going $\nutau$ signal to be a competitive search.
A possible way to reduce the misidentification background could be
obtained by considering specific
tau hadronic decays.
Indeed, the tau lepton decays mainly producing multiple pion events: about 40\%
of the hadronic
decays are given by $\tau^- \rightarrow \pi^0 \pi^- \nu_\tau$.
This channel produces two electromagnetic cascades from
$\pi^0\rightarrow\gamma \gamma$
decay and one muon track from $\pi^- \rightarrow \mu^- \bar{\nu}_\mu$ decay. 
The main experimental challenge would be to find suitable cuts to
statistically
distinguish these events from the CC/NC $\nu_e$ and $\nu_\mu$ multiple
pion productions.

\subsection{\label{sec:EventBKGCC} Background events from $\nue$ charged--current interactions}

Another source of background for the hadronic tau decay is represented by 
the charged--current $\nu_e$ channel. 
We calculate background events due to $\nu_e$ charge current 
interactions as \cite{Giordano:2010pr}:

\be
\left.N^{\mathcal{CC}}_{e}\right|_\mathcal{B}=M_{\rm det} N_y \times 
\int^{E^{\rm max}_{\rm vis}}_{E^{\rm min}_{\rm vis}}
dE_{\rm vis}\,
\int d \Omega\;\;\eta(\theta)\,\left. \frac{d \phi_{\nu_e}}{d \Omega d E_{\rm vis}} \right|_\mathcal{B}
\Sigma^{\mathcal{CC}}_{e,{\rm TOT}}(E_{\rm vis})
+(\nu \rightarrow \bar{\nu})\,,
\ee
where $\Sigma^\mathcal{CC}_{e,{\rm TOT}}$ is the analogous of Eq.~\eqref{eq:sigmaCC} for 
the total charged--current cross section of $\nu_e$, 
for which we used the deep--inelastic expressions of Ref.~\cite{Strumia:2006db}. 
Note that, in this case, the visible energy $E_{\rm vis}$ is equal to the full initial 
neutrino energy $E_\nu$.

The number of charged current $\nu_e$ events for one 
Mton$\times$year exposure is reported again in Fig.~\ref{fig:events}. We
notice that this is the category of events which (through misidentification)
represents the largest source of background.

\section{\label{sec:SS} Detectability and statistical significance}

To quantify the discovery reach of present and future water Cherenkov 
detectors, we use the statistical significance $\varsigma$, defined 
as the signal--to--noise ratio: 

\be
\varsigma\equiv\frac{S}{\sqrt{S+B}}\,.
\ee
See also Ref.~\cite{Zhu:2008yh} for a complete discussion on the statistical significance. 
We have studied the behaviour of $\varsigma$ in two cases: an ideal 
case in which no misidentification is present and the detector efficiency 
for tau leptons is almost 100\%, and a more realistic case in which both 
the misidentification and the detection efficiency for taus are considered.
In the ideal case we have: 
\be
B_{\rm ideal}=\left.N^\mathcal{CC}_\tau\right|_\mathcal{B}\,.
\ee
In the presence of misidentification of the electron 
$\epsilon^{\rm mis}_e$ and muon $\epsilon^{\rm mis}_\mu$ events
and detection efficiency $\epsilon_\tau$ for the detection of tau hadronic events, we instead have: 

\be
B_{\rm realistic}=
\epsilon_\tau \left.N^\mathcal{CC}_\tau\right|_\mathcal{B} \,+ \,
\epsilon^{\rm mis}_\mu \left.N^\mathcal{NC}_\mu\right|_\mathcal{B} \,+ \,
\epsilon^{\rm mis}_e \left(\left.N^\mathcal{NC}_e\right|_\mathcal{B} + 
\left.N^\mathcal{CC}_e\right|_\mathcal{B}\right)\,.
\ee

\begin{figure}[t]
\includegraphics[width=0.96\textwidth]{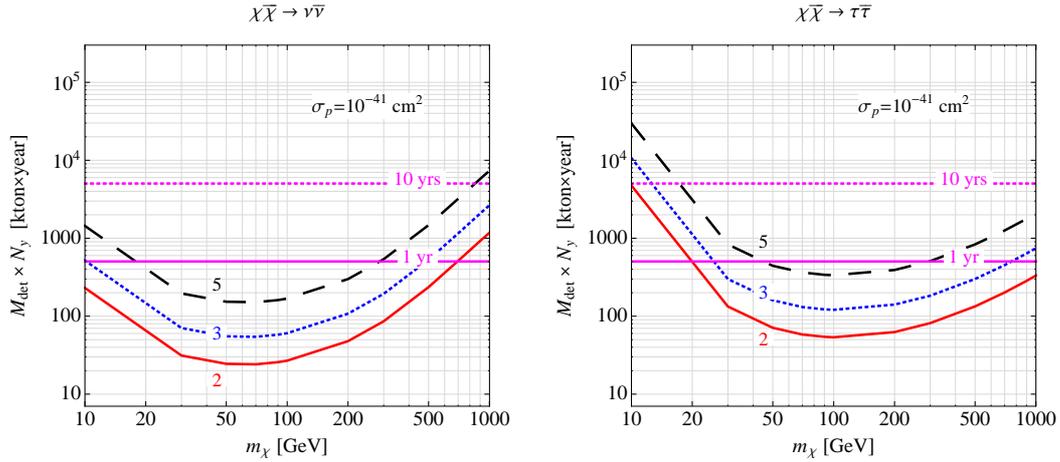}
\caption
{\label{fig:Exposure_significance} 
Iso--contours of statistical significance for the detection of
downward--going $\nutau$ hadronic events 
in the plane of detector exposure (in kton $\times$ year) and DM
mass $m_\chi$. The left panel refers to DM
annihilation into neutrinos, while the right panels
stands for annihilation into tau leptons. The elastic scattering cross section
on protons (relevant for capture in the Sun) is fixed at the benchmark value of $10^{-41}$ cm$^{2}$.
The solid (red), dotted (blue)
and dashed (black) lines refer to significance of 2, 3 and 5~$\sigma$.
The two horizontal (pink) lines denote the exposures that can be reached
by a 0.5 Mton detector, like HK, in 1 (solid) and 10 (dotted) years. In this plot
we assumed full efficiency for the detection of tau hadronic events ($\epsilon_\tau = 100\%$) and no 
misidentification of electron and muon NC/CC events 
($\epsilon^{\rm mis}_e = \epsilon^{\rm mis}_\mu = 0$).}
\end{figure}

\begin{figure}[t]
\includegraphics[width=0.96\textwidth]{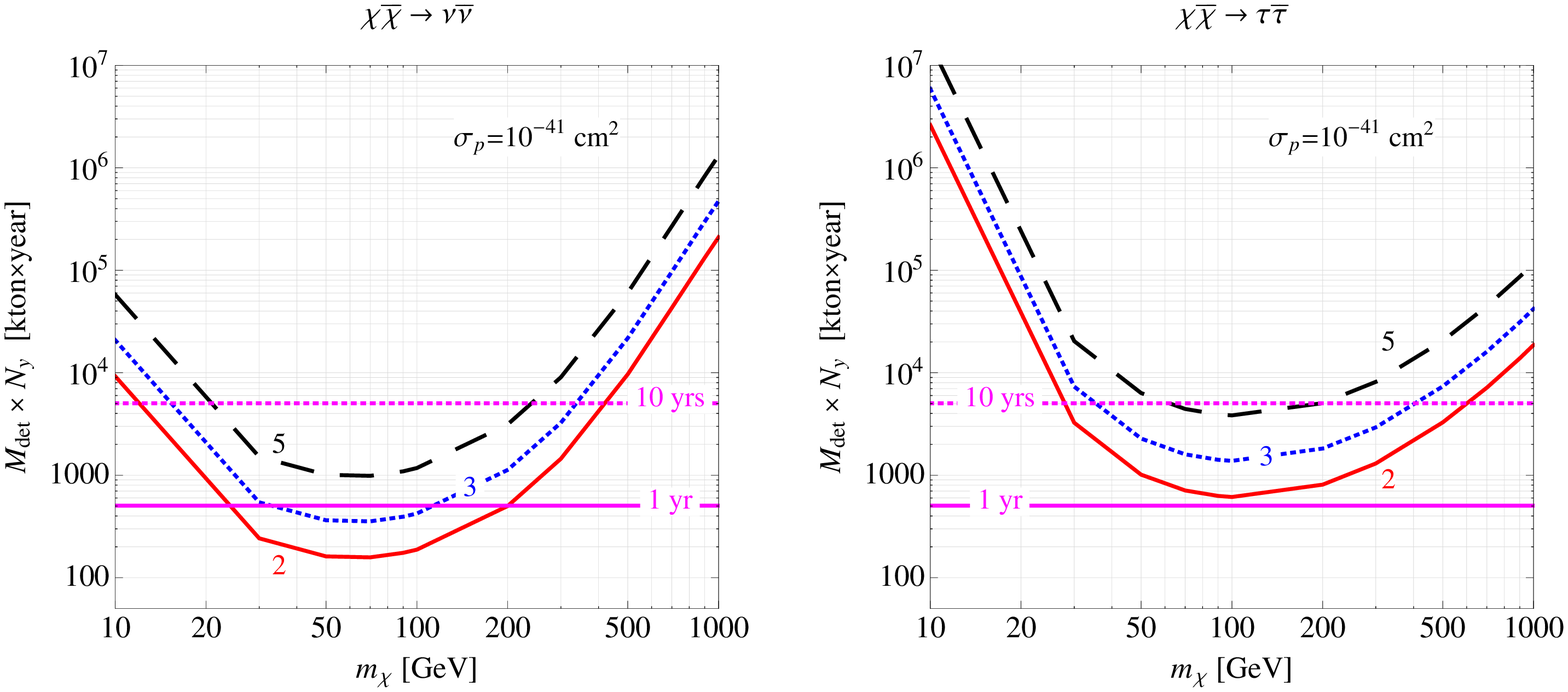}
\caption
{\label{fig:Exposure_significance_MIS} 
The same as in Fig.~\ref{fig:Exposure_significance}, for 
detection efficiency of tau hadronic events $\epsilon_\tau = 40\%$ and misidentification of electron 
and muon events $\epsilon^{\rm mis}_e = \epsilon^{\rm mis}_\mu = 4\%$.}
\end{figure}

\begin{figure}[t]
\includegraphics[width=0.96\textwidth]{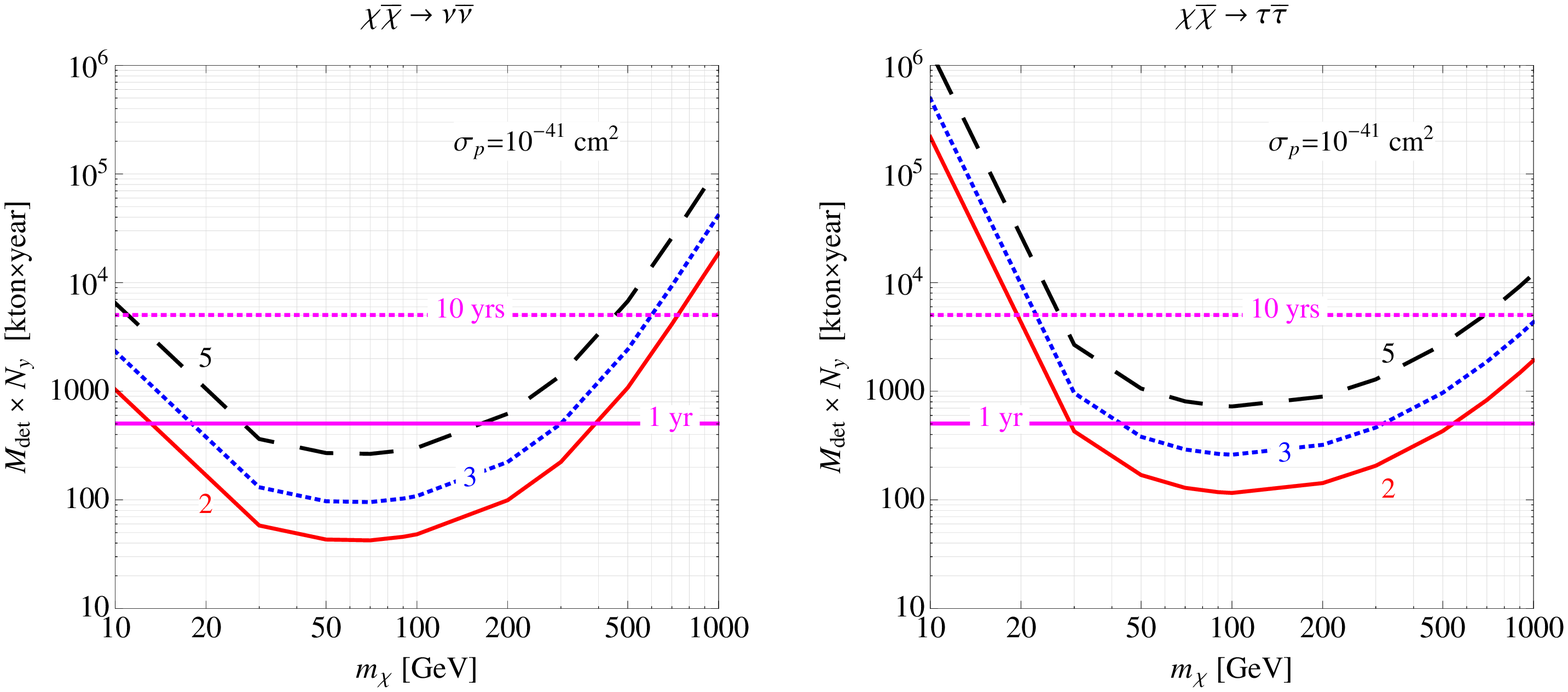}
\caption
{\label{fig:Exposure_significance_MIS_opt} 
The same as in Fig.~\ref{fig:Exposure_significance}, for 
detection efficiency of tau hadronic events $\epsilon_\tau = 70\%$ and misidentification of electron 
and muon events $\epsilon^{\rm mis}_e = \epsilon^{\rm mis}_\mu = 1\%$.}
\end{figure}

Our results are reported in Fig.~\ref{fig:Exposure_significance},
\ref{fig:Exposure_significance_MIS} and \ref{fig:Exposure_significance_MIS_opt}.
Fig.~\ref{fig:Exposure_significance} shows the
iso--contours of statistical significance for the detection of
downward--going $\nutau$ hadronic events 
in the plane of detector exposure (in kton $\times$ year) and DM
mass $m_\chi$. The left panel refers to DM
annihilation into neutrinos, while the right panels
stands for annihilation into tau leptons. The elastic scattering cross section
on protons (relevant for capture in the Sun) is fixed at the benchmark value of $10^{-41}$ cm$^{2}$.
The solid (red), dotted (blue)
and dashed (black) lines refer to significance of 2, 3 and 5~$\sigma$.
The two horizontal (pink) lines denote the exposures that can be reached
by a 0.5 Mton detector, like HK, in 1 (solid) and 10 (dotted) years. This plot
refers to an ideal detector with full efficiency for the detection of tau hadronic 
events ($\epsilon_\tau = 100\%$) 
respect to the neutral and charged current electron and muon events
($\epsilon^{\rm mis}_e = \epsilon^{\rm mis}_\mu = 0$). We notice that in this
ideal case a 5~$\sigma$ significance of signal detection is present for a wide
DM mass range (from about 20 to 300 GeV) with 1 year of exposure of 1 Mton
detector. Indication at the 2--3~$\sigma$ level
are possible up to 500-700 GeV for the case of direct annihilation into neutrinos,
while it extends to larger masses (and drops al small DM masses) for the 
case of annihilation into tau leptons. Clearly, statistical significance scales with
the scattering cross section $\sigma_p$. 
Note that a more precise analysis would require to take into account the 
theoretical uncertainties on the intrinsic $\nu_\tau$ fluxes~\cite{Costa:2001fb,*Costa:2001va,Costa:2000jw,Enberg:2008te}. 
The intrinsic contribution constitutes roughly 21\% of the total $\nu_\tau$ events calculated 
as described in Sect.~\ref{sec:EventDM}. 
For an intrinsic flux one order of magnitude higher than the one considered here~\cite{Pasquali:1998xf}, the curves for the statistical significance 
$\varsigma$ in Fig.~\ref{fig:Exposure_significance} 
would move up by a factor of 1.3 for the $\tau\bar{\tau}$ annihilation channel 
and for a DM mass 
$m_\chi$ between 10 and 20 GeV. The curves will remain roughly unchanged for 
higher masses and for the $\nu\bar{\nu}$ annihilation channel. 
We will moreover see that in the realistic cases of our interest (with misidentification) 
the dependence on the intrinsic $\nu_\tau$ flux is highly reduced. 

A more realistic situation requires to take under proper consideration the 
efficiency for reconstruction of tau events and the
misidentification of the $\nu_e$ and $\nu_\mu$ events. 
This is shown in Fig.~\ref{fig:Exposure_significance_MIS}, where we reduce the efficiency 
to $\epsilon_\tau = 40\%$ and we allow a misidentification of 4\%, both
in the electron and muon channels, not far from those already achieved by the SK 
analysis~\cite{Abe:2006fu,Kato:2007zz}. 
We notice that now large statistical significance
is recovered with a much larger exposure: nevertheless, a 10 yr period of
data taking reproduces the same level of reach as the one discussed for the optimal
case in Fig.~\ref{fig:Exposure_significance}. Fig.~\ref{fig:Exposure_significance_MIS_opt}
shows the capabilities in the case of better performance:
$\epsilon_\tau = 70\%$ and $\epsilon^{\rm mis}_e = \epsilon^{\rm mis}_\mu = 1\%$.
In this case, a few years of exposure would suffice to cover almost 
the whole DM mass range (with noticeable differences for different annihilation channels)
for our benchmark value of the scattering cross section $\sigma_p$. 
Note that also for $\epsilon^{\rm mis}_e = \epsilon^{\rm mis}_\mu = 1\%$ 
the background is still dominated by misidentified events, see Fig.~\ref{fig:events}, and the uncertainties on 
the $\nu_\tau$ intrinsic flux (which account for roughly  
21\% of the total number of events) does not alter our conclusions.

\begin{figure}[t]
\includegraphics[width=0.96\textwidth]{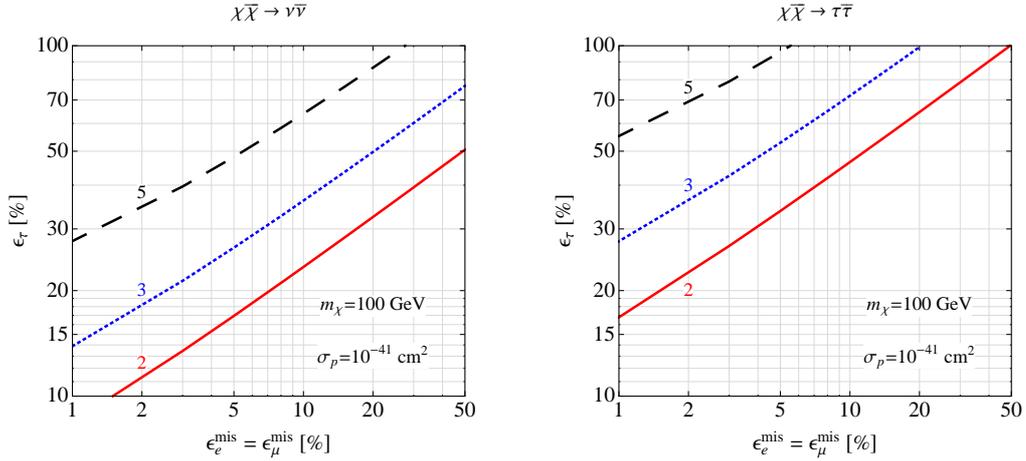}
\caption
{\label{fig:Epsilon_MIS} 
Dependence of the statistical significance $\varsigma$ on the misidentification parameters 
$\epsilon_e=\epsilon_\mu$ and $\epsilon_\tau$, for a DM mass $m_\chi=100$~GeV with 
cross section $\sigma_p=10^{-41} {\rm cm}^2$ and 
an exposure $M_{\rm det} N_y=1$ Mton$\times$year. The solid (red), dotted (blue)
and dashed (black) lines refer to $\varsigma = 2,3,5$. The left panel stands for DM annihilation 
into neutrinos, while the right panels
for annihilation into tau leptons.}
\end{figure}

The dependence
of the statistical significance with the detection efficiencies is reported,
as an illustrative example, in
Fig.~\ref{fig:Epsilon_MIS} for a DM mass $m_\chi=100$ GeV and the benchmark value
of $\sigma_p$. We notice that an efficiency of tau--events reconstruction of
50--60\% would allow to a clear signal detection even without the need to reduce
the level of misidentification. At the same time, an alternatively, a reduction
of $\epsilon^{\rm mis}_e$ and $\epsilon^{\rm mis}_\mu$ at the 1\% level would be
enough (for this benchmark case) to obtain a 5~$\sigma$ detection quite easily.

\begin{figure}[t]
\centering
\includegraphics[width=0.46\textwidth]{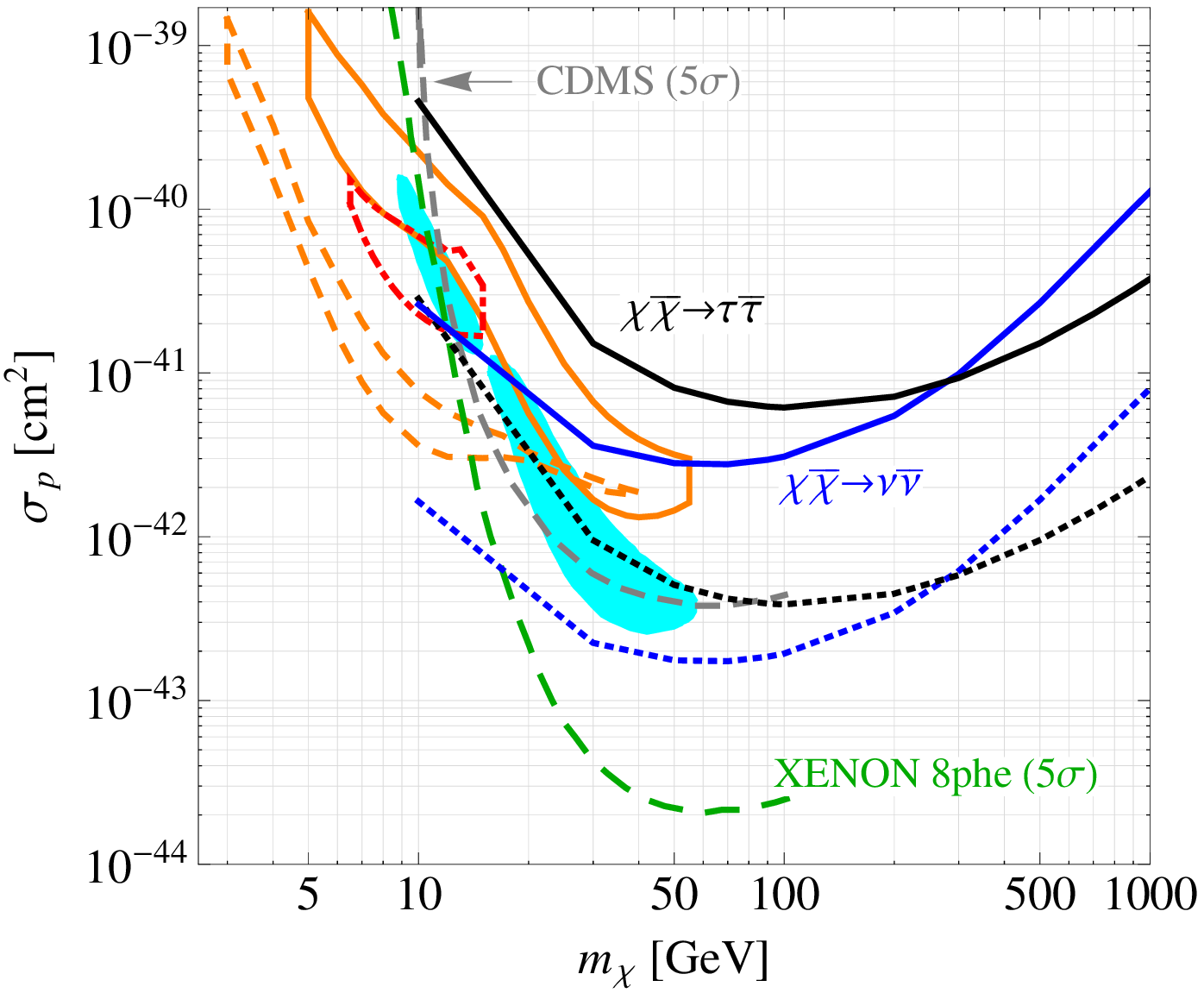}
\includegraphics[width=0.46\textwidth]{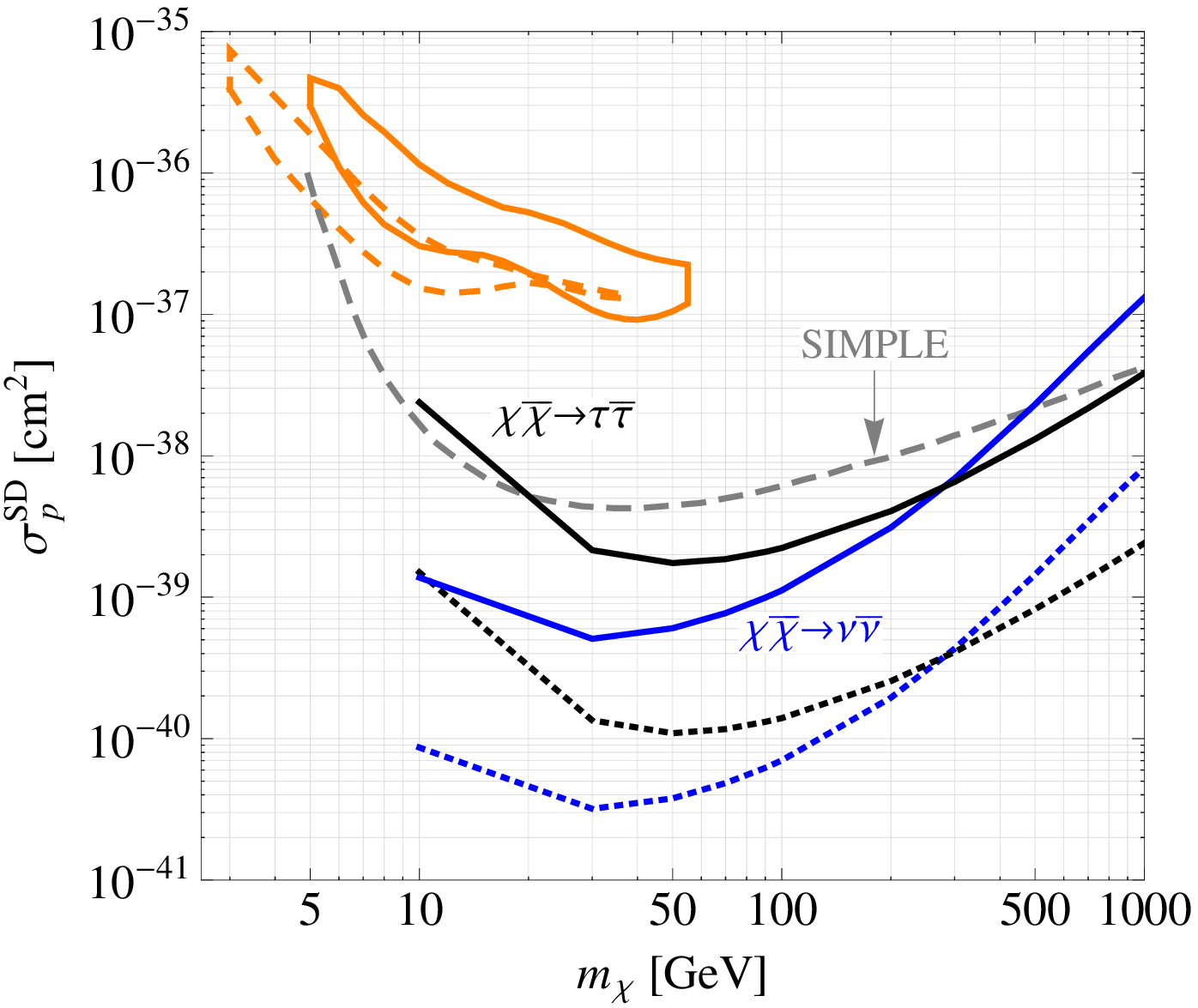}
\caption
{\label{fig:SigmaP_mChi} 
Dependence of the statistical significance $\varsigma$
on the DM scattering cross section on protons $\sigma_p$
as a function of the DM mass $m_\chi$, that can be derived
by using the downward--going hadronic events with an exposure 
$M_{\rm det} N_y=$1~Mton$\times$year. The curves refer to $\varsigma = 1.64$ 
and can alternatively be considered as the 90\% C.L. upper bounds deriving 
from the downward--going hadronic events on $\sigma_p$ vs. $m_\chi$. 
The dotted lines represent the limits 
without considering misidentification, while the solid lines 
refer to $\epsilon_\tau$=40\% and
$\epsilon^{\rm mis}_e$=$\epsilon^{\rm mis}_\mu$=4\%. 
Left panel: our results for the spin independent case 
together with the allowed regions 
from DAMA (orange solid line for the case without channeling, 
orange dashed line for the channeling case), CoGeNT (dot--dashed red curve) and CRESST (cyan regions), and the limits from XENON 100 (green dashed line) 
and CDMS (gray dashed line) experiments. 
Right panel: our results for the spin--dependent case, together with 
the allowed regions from DAMA (upper for no--channeling, lower for channeling) and the limit from SIMPLE (gray dashed line).}
\end{figure}

In order to test the sensitivity of the downward--going tau signal not only to the 
DM mass, but also to its scattering cross section $\sigma_p$, we show in 
Fig.~\ref{fig:SigmaP_mChi} the contours for $\varsigma = 1.64$ (which corresponds
to a C.L. of 90\%) in the plane $\sigma_p$ vs. $m_\chi$, for a detector with
exposure $M_{\rm det} N_y=$1~Mton$\times$year.
The dotted lines represent the limits 
without considering misidentification, while the solid lines 
refer to $\epsilon_\tau$=40\% and
$\epsilon^{\rm mis}_e$=$\epsilon^{\rm mis}_\mu$=4\%. The left 
panel refers to spin independent interactions, while 
the right panel shows the case for a spin dependent cross section. 
These curves can
alternatively be considered as the 90\% C.L. upper bounds deriving 
from the downward--going hadronic events on $\sigma_p$ vs. $m_\chi$. 
In the left panel of Fig.~\ref{fig:SigmaP_mChi}, beside our limits, we show the allowed regions obtained considering the DAMA~\cite{Bernabei:1998fta,Bernabei:2008yi,Bernabei:2010mq}, 
CoGeNT~\cite{Aalseth:2011wp} and CRESST~\cite{Angloher:2011uu} positive results. 
The DAMA and CoGeNT regions are taken from Ref.~\cite{Belli:2011kw} to which we refer 
for more detailed information. 
The DAMA region represents the domain where the likelihood function 
values differ more than 7.5~$\sigma$ from the null hypothesis (absence of modulation), 
while the CoGeNT region refers to the area where the likelihood function 
values differ more than 1.64~$\sigma$. 
We also plot the constraints from XENON~\cite{Aprile:2011hi} and CDMS~\cite{Ahmed:2009zw} 
experiments, as calculated in the statistical analysis of Ref.~\cite{Fornengo:2011sz}. 
These limits are taken at 5~$\sigma$ and the XENON threshold has been set to 8 photoelectrons. 
In the right panel of Fig.~\ref{fig:SigmaP_mChi}, we report the DAMA region in the 
case of spin-dependent interaction~\cite{DAMAprivate}. We also report the limit on
spin--dependent 
interactions from the SIMPLE experiment~\cite{Felizardo:2011uw}, see also Ref.~\cite{Collar:2011kr} and Ref.~\cite{Collaboration:2011ig} for discussions on critical points.
Further weaker limits, not reported in Fig.~\ref{fig:SigmaP_mChi}, come from the PICASSO~\cite{Archambault:2009sm}, COUPP~\cite{Behnke:2010xt} 
and KIMS~\cite{Lee.:2007qn} experiments.
The recent analyses of direct 
detection annual modulation effects observed by DAMA~\cite{Bernabei:1998fta,Bernabei:2008yi,Bernabei:2010mq}
and CoGeNT~\cite{Aalseth:2011wp}
(and the excess of events reported by CRESST~\cite{Angloher:2011uu}) point toward a light
DM candidate with a mass around 10 GeV and spin--independent
scattering cross sections of the order of 10$^{-42}$ cm$^2$ -- 10$^{-40}$ cm$^2$. For this 
type of particle, we would expect, for the case of direct annihilation into neutrinos,
 between 9 and 900 hadronic events and a detection with a statistical
significance close to 5~$\sigma$ with a 10 years exposure on HK (5 Mton$\times$yr)
in the case of the improved performance of Fig.~\ref{fig:Exposure_significance_MIS_opt}. 


\section{\label{sec:conclusions} Conclusions}

In the context of indirect DM searches with neutrinos, the most common channel
of investigation is represented by upward--going muons or by contained $\mu$-like 
or $e$-like 
events produced by the charged--current conversion of the muon-- or 
electron--neutrino fluxes produced by DM annihilation
in the Sun or in the Earth. These channels are very solid and do not suffer from 
large detection difficulties. In the most typical case of upward--going muons, the
conversion area for the $\numu \rightarrow \mu$ process is represented by a large
portion of the rock below the detector and experimental apparata posses very large detection
efficiencies. For instance, SK has almost a 100\% efficiency 
to detect through--going muons. 
The most important limit for this type of DM searches, instead, is
represented by the large $\numu$ and $\nue$ atmospheric background that cast a shadow on the DM signal. 

In this paper we propose a new channel for DM
searches at
neutrino telescopes: downward--going hadronic tau events
originated by the $\nutau$ signal produced by DM annihilation in the Sun.
This specific signal potentially represents a very good opportunity for
DM, since the background of atmospheric downward--going $\nutau$ is 
extremely reduced with respect to the upward--going $\numu$ case commonly considered.
The intrinsic amount of $\nutau$ in atmospheric neutrinos is very small as compared
to $\nue$ and $\numu$ components, while the $\nutau$ component in the signal
from DM annihilation (or decay) in the Sun is typically expected to be of the
same order of their $\nue$ and $\numu$ counterparts. Moreover, in the downward--going direction, atmospheric
$\numu$ do not have enough baseline to oscillated into $\nutau$. Therefore, a flux of tau neutrinos coming from 
the Sun when the star is above the horizon represents a signal with 
a very reduced background. The signal--to--background ratio in terms of
fluxes is therefore much more favorable for down--going $\nutau$'s then for 
up--going $\numu$'s and $\nue$'s (or even $\nutau$'s
themselves).

Additional sources of background are  represented by $\nutau$ produced in the 
solar corona of the Sun (which represent an irreducible background for our DM signal, 
since it comes from the same source, the Sun)
and in $\nutau$ arriving at the Earth from the galactic
plane, mainly produced 
by oscillation of $\numu$ produced in cosmic--ray interactions with the interstellar 
medium. 
This source of background is, in 
principle, reducible by angular cuts on the galactic plane position. For detectors 
located in the northern hemisphere the angles $\cos \theta_Z \geq 0.5$ are particularly 
favourable, because the galactic plane duty--factor is small for those angles. 
We have shown
that all these sources of backgrounds are, by themselves, not posing relevant limits
to the DM signal, which therefore represent a potentially viable novel possibility.

The large advantage represented by the reduced background for the downward--going
$\nutau$ fluxes is, unfortunately, partly diminished by  
limits inherent in the difficulty to detect and properly identify tau neutrinos.
Indeed, detectors specifically designed for the identification of tau events, like 
Opera at Gran Sasso Laboratory, are currently too small to collect a statistically 
significant number of events. 
For this reason we have focused our analysis on Cherenkov detectors. 
In this case the hadronic tau events cannot be easily distinguished by 
the neutral current events, mostly coming from atmospheric $\nu_e$ and $\nu_\mu$, 
and by the charged current $e$--like events. 
The possibility to correctly detect hadronic tau events is currently based on 
statistical methods and the percentage of misidentified events for a SK--like 
detector is of the order of several percent \cite{Abe:2006fu,Kato:2007zz}. As the 
atmospheric $\nu_e$ and $\nu_\mu$ 
are more abundant than the signal $\nu_\tau$ events, the misidentification influences 
the discovery potential of DM through the downward--going tau channel. 

Since for existing Cherenkov detectors, like SK, the 
number of hadronic tau events expected is small, we have focused our analysis
on future Mton--size Cherenkov 
detectors, like Hyper--Kamiokande~\cite{Nakamura:2003hk}, 
UNO~\cite{Jung:1999jq} or 
MEMPHYS~\cite{deBellefon:2006vq}, respectively in Japan, US and France. 
All these future detectors will be situated in the norther hemisphere and this 
could represent a great advantage to reduce the background of galactic neutrinos 
in the study of events from zenith angles $\cos \theta_Z \geq 0.5$. 
We have not discussed in our analysis the IceCube detector, since 
$\nutau$ reconstruction, at the low energies required to study the DM 
in the mass range of interest in our analysis (10 GeV -- 1 TeV), may not be possible, but 
the possibility to implement statistical analyses is under study~\cite{rott:2009}.
For the detection of high energy $\nu_\tau$ at IceCube 
see for instance Ref.~\cite{Seo:2007zzd}. 

For the Mton--size water Cherenkov detectors we have shown that the
downward--going tau neutrinos signal has potentially  good prospects, the main limitation being the level of 
misidentification of
non--tau events, which need to be kept at the level of percent. For
definiteness, we have studied two benchmark cases: DM directly
annihilating into neutrinos, with equal amount of the three active flavors; dark
matter annihilating into tau pairs. We showed that several tens of
events per year (depending on the DM mass and annihilation/decay channel) are potentially collectible in 
a Mton--scale detector. Once the misidentification
of non--tau events is taken under consideration, a 5~$\sigma$ significance discovery is potentially reachable for 
DM masses in the range from 20 to 300 GeV with a few years of exposure, and for a benchmark value of DM 
scattering cross section on protons (relevant for DM capture in the Sun) of $\sigma_p = 10^{-41}$ cm$^{-2}$.
For light DM candidates with a mass around 10 GeV and spin--independent
scattering cross sections of the order of 10$^{-42}$ cm$^2$ -- 10$^{-40}$ cm$^2$,
which are of special current interest due to the recent results in direct detection
studies, we find that, for the case of direct annihilation into neutrinos,
a detector like HK within 10 years (exposure of 5 Mton$\times$yr) could collect
between 9 and 900 hadronic tau DM events and be sensitive to 
detection with a statistical significance close to 5~$\sigma$, in the case
of a 70\% detection efficiency in the reconstruction of the tau hadronic events and
of a 1\% level of misidentification of non--tau events.

\acknowledgments

We would like to thank G. Giordano, M.C. Gonzalez--Garcia, O. Mena 
and I. Mocioiu for useful discussions. 
We acknowledge Research Grants funded jointly by Ministero
dell'Istruzione, dell'Universit\`a e della Ricerca (MIUR), by
Universit\`a di Torino and by Istituto Nazionale di Fisica Nucleare
within the {\sl Astroparticle Physics Project} (MIUR contract number: PRIN 2008NR3EBK;
INFN grant code: FA51). N.F. acknowledges support of the spanish MICINN
Consolider Ingenio 2010 Programme under grant MULTIDARK CSD2009- 00064.
This work was partly completed at the Theory Division of CERN in the context of the TH--Institute 
`Dark Matter Underground and in the Heavens' (DMUH11, 18-29 July 2011).


\bibliographystyle{JHEP}
\bibliography{nu_tau}


\end{document}